\documentclass[prb,aps,twocolumn,showpacs]{revtex4}
\usepackage{graphicx}

\begin{document}

\title{       Theory of optical spectral weights in Mott insulators 
              with orbital degrees of freedom 
}

\author {     Giniyat Khaliullin }
\affiliation{ Max-Planck-Institut f\"ur Festk\"orperforschung,
              Heisenbergstrasse 1, D-70569 Stuttgart, Germany \\ and
              E. K. Zavoisky Physical-Technical Institute of the
              Russian Academy of Sciences, 420029 Kazan, Russia } 
\author {     Peter Horsch }
\affiliation{ Max-Planck-Institut f\"ur Festk\"orperforschung,
              Heisenbergstrasse 1, D-70569 Stuttgart, Germany }
\author {     Andrzej M. Ole\'{s} }
\affiliation{ Max-Planck-Institut f\"ur Festk\"orperforschung, 
	      Heisenbergstrasse 1, D-70569 Stuttgart, Germany \\ 
	      and Marian Smoluchowski Institute of Physics, Jagellonian
              University, Reymonta 4, PL-30059 Krak\'ow, Poland}

\date{9 June 2004}
%\date{17 March 2004}
%\date{9 January 2004}
%\date{17 September 2003}

\begin{abstract}
Introducing partial sum rules for the optical multiplet transitions,
we outline a unified approach to magnetic and optical properties of 
strongly correlated transition metal oxides. On the example of LaVO$_3$ 
we demonstrate how the temperature and polarization dependences of 
different components of the optical multiplet are determined by the 
underlying spin and orbital correlations dictated by the low-energy 
superexchange Hamiltonian. Thereby the optical data provides deep 
insight into the complex spin-orbital physics and the role played by 
orbital fluctuations.
\end{abstract}


\maketitle

\section{Introduction}

%%%%%%%%%%%%%%%%%%%%%%%%%%%%%%%%%%%%%%%%%%%%%%%%%%%%%%%%%%%%%%%%%%%%%%%%
%%
%%%%%%%%%%%%%%%%%%%%%%%%%%%%%%%%%%%%%%%%%%%%%%%%%%%%%%%%%%%%%%%%%%%%%%%%
Charge localization in Mott insulators is not perfect as electrons still 
undergo virtual transitions to neighboring sites in order to retain 
partially their kinetic energy. These high-energy virtual transitions 
across the Mott-Hubbard gap $\sim U$ are crucial for magnetism --- 
this quantum charge motion leads to superexchange (SE) interactions 
\cite{And59} between local degrees of freedom. It is frequently not 
realized, however, that the same charge excitations are responsible in 
Mott insulators for the low-energy optical absorption. Therefore, the 
intensity of the optical absorption and the SE energy are intimately 
related to each other via the well known optical sum rule,\cite{Bae86} 
which links the integrated optical conductivity and the kinetic energy. 
In Mott insulators the latter is due to virtual exchange processes and 
hence the thermal evolution of spectral weights and SE energy are 
related.\cite{Aic02} 

%%%%%%%%%%%%%%%%%%%%%%%%%%%%%%%%%%%%%%%%%%%%%%%%%%%%%%%%%%%%%%%%%%%%%%%%
%%
%%%%%%%%%%%%%%%%%%%%%%%%%%%%%%%%%%%%%%%%%%%%%%%%%%%%%%%%%%%%%%%%%%%%%%%%
A qualitatively new situation is encountered in Mott insulators with
partly filled $d$ orbitals. Because of orbital degeneracy, virtual 
charge excitations that are seen in optics reflect the rich multiplet 
structure of a transition metal ion determined by Hund's exchange 
coupling $J_H$, and orbital degrees of freedom contribute then to the 
SE.\cite{Kug82} When spin and orbital correlations change the individual 
components of the optical multiplet reflect characteristic {\it spectral 
weight transfer\/}. Moreover, the cubic symmetry is spontaneously broken 
by orbital and spin order, and thus one expects {\it anisotropic optical 
absorption}. Indeed, pronounced anisotropy was reported for LaMnO$_3$, 
\cite{Tob01} both for the $A$-type antiferromagnetic (AF) phase 
\cite{noteaaf} and for the orbital ordered phase above the N\'eel 
temperature $T_N$. Recently, the anisotropy in optical absorption and 
its strong temperature dependence near the magnetic transitions were 
found for cubic vanadates.\cite{Miy02} This latter example is even more 
puzzling as the magnetic properties are anomalous,\cite{Mah92} and 
neutron scattering experiments \cite{Ulr03} have revealed nontrivial 
quasi one-dimensional (1D) correlations of spin and orbital degrees of 
freedom that are surprising for crystals with nearly cubic symmetry. 
Indeed, a theory of spin and orbital states in cubic vanadates predicted
quasi 1D spin-orbital correlations due to a spontaneous breaking of the 
cubic symmetry in the SE model.\cite{Kha01}

%%%%%%%%%%%%%%%%%%%%%%%%%%%%%%%%%%%%%%%%%%%%%%%%%%%%%%%%%%%%%%%%%%%%%%%%
%%
%%%%%%%%%%%%%%%%%%%%%%%%%%%%%%%%%%%%%%%%%%%%%%%%%%%%%%%%%%%%%%%%%%%%%%%%
It is our aim to outline a unified picture that links optical and 
magnetic properties at orbital degeneracy. Starting from the low-energy 
spin-orbital model we derive {\it partial sum rules} for the different 
excited states. Thereby we provide a rigorous theoretical basis for
the analysis of optical spectral weights and show how the evolution of 
magnetic coherence manifests itself in optics as {\it intensity 
transfer\/} between different excitations (upper Hubbard bands). 
This explains the origin of dramatic variation of the optical absorption 
and its anisotropy with temperature $T$ in manganites,\cite{Tob01} 
vanadates,\cite{Miy02} and ruthenates,\cite{Lee02} where at low $T$ 
the high-spin band carries most spectral weight for the directions with 
ferromagnetic (FM) spin correlations. We illustrate this idea for the 
case of $C$-type AF ($C$-AF) phase \cite{noteaaf} of cubic vanadates 
with degenerate $t_{2g}$ orbitals. We show that the predicted quasi 1D 
spin-orbital correlations,\cite{Kha01} realized in $C$-AF phase of 
LaVO$_3$, are reflected in the $T$-dependence of optical weights 
derived from the SE model.

%%%%%%%%%%%%%%%%%%%%%%%%%%%%%%%%%%%%%%%%%%%%%%%%%%%%%%%%%%%%%%%%%%%%%%%%
%%
%%%%%%%%%%%%%%%%%%%%%%%%%%%%%%%%%%%%%%%%%%%%%%%%%%%%%%%%%%%%%%%%%%%%%%%%
The paper is organized as follows. First we present a generic structure 
of the SE interactions in a Mott insulator with orbital degrees of 
freedom in Sec. II, and relate them to the intensities in optical 
absorption. On the example of the SE interactions encountered in 
LaVO$_3$, we analyze next spin, orbital, and joint spin-and-orbital 
correlations which determine the optical intensities. In this way, we 
arrive at a set of self-consistent equations which are solved in Sec. 
III, where we present the numerical results for the optical sum rules 
for individual high-spin and low-spin excitations, as well as for the 
spin and orbital SE interactions. These results and their comparison 
with experiment are discussed in Sec. IV, where we also give a summary 
and more general conclusions.

\section{Theory}

%%%%%%%%%%%%%%%%%%%%%%%%%%%%%%%%%%%%%%%%%%%%%%%%%%%%%%%%%%%%%%%%%%%%%%%%
%%
%%%%%%%%%%%%%%%%%%%%%%%%%%%%%%%%%%%%%%%%%%%%%%%%%%%%%%%%%%%%%%%%%%%%%%%%
The SE interaction in a cubic Mott insulator with ions  
having orbital degrees of freedom has a generic form, 
\begin{equation}
\label{sex}
{\cal H}_{J}=H_s+H_{\tau}+H_{s\tau}=
\sum_n\sum_{\langle ij\rangle\parallel\gamma} H_n^{(\gamma)}(ij),   
\end{equation}
and consists of separate spin ($H_s$) and orbital ($H_{\tau}$) 
interactions, and of a dynamical coupling between them ($H_{s\tau}$). 
This complex form of ${\cal H}_J$, given by Eq. (\ref{sex}), follows 
from the terms $H_n^{(\gamma)}(ij)$ for each bond $\langle ij\rangle$
along a given cubic axis $\gamma=a,b,c$, 
arising from the transitions to various upper Hubbard bands labelled 
by $n$. The optical intensity of {\it each band\/} $n$, for the photon 
polarization along a cubic axis $\gamma$, is determined by the 
respective SE energy:
\begin{equation}
\label{central}
\frac{a_0\hbar^2}{e^2}\int_0^{\infty}\sigma_n^{(\gamma)}(\omega)d\omega=
-\frac{\pi}{2}K_n^{(\gamma)}=
-\pi\Big\langle H_n^{(\gamma)}(ij)\Big\rangle.
\end{equation}
Here, $a_0$ is the distance between magnetic ions (the tight-binding
model is implied), and $\big\langle H_n^{(\gamma)}(ij)\big\rangle$ is
the SE interaction for a bond $\langle ij\rangle$ along axis $\gamma$. 
The first equality in Eq. (\ref{central}) follows from the optical sum 
rule for a given transition $n$, and relates the kinetic energy 
$K_n^{(\gamma)}$ to the optical conductivity 
$\sigma_n^{(\gamma)}(\omega)$ for this band, while the second equality 
relates the associated kinetic energy to the SE energy via the 
{\it Hellman-Feynman theorem\/}.\cite{Bae86}

Experimental data is often presented in terms of an effective carrier
number [see, e.g., Eq. (2) of Ref. \onlinecite{Miy02}], 
$N_{{\rm eff},n}^{(\gamma)}=(2m_0v_0/\pi e^2)\int_0^{\infty}
\sigma_n^{(\gamma)}(\omega)d\omega$, where $m_0$ is the free electron 
mass, and $v_0=a_0^3$ is the volume per magnetic ion. This gives an 
{\it optical sum rule\/} as follows:
\begin{equation}
\label{sumrule}
N_{{\rm eff},n}^{(\gamma)}=-\frac{m_0a_0^2}{\hbar^2}\;K_n^{(\gamma)}
=-\frac{m_0a_0^2}{\hbar^2}\;\Big\langle 2H_n^{(\gamma)}(ij)\Big\rangle.
\end{equation}
Each level $n$ of the multiplet represents an upper Hubbard band with 
its own spin and orbital quantum numbers. The key point is that while  
full kinetic energy and corresponding total intensity may show only 
modest $T$-dependence and almost no anisotropy, the behavior of the 
individual transitions is much richer, and directly reflects the 
ground state correlations via the spin and orbital selection rules. 

%%%%%%%%%%%%%%%%%%%%%%%%%%%%%%%%%%%%%%%%%%%%%%%%%%%%%%%%%%%%%%%%%%%%%%%%
%%
%%%%%%%%%%%%%%%%%%%%%%%%%%%%%%%%%%%%%%%%%%%%%%%%%%%%%%%%%%%%%%%%%%%%%%%%
Hund's exchange separates the lowest (high-spin) excitation from the 
next (low-spin) one by: $\sim 5J_H$ in manganites,\cite{Fei99} 3$J_H$ 
in vanadates,\cite{Kha01} and 2$J_H$ in titanates.\cite{Kha00} When 
the high-spin excitation, broadened by its propagation in crystal and 
by many-body effects, has a smaller linewidth than the above multiplet 
splitting, it may show up in optical spectroscopy as a separate band. 
This is in fact nearly satisfied for typical values of 
$J_H\sim 0.6-0.7$ eV and hoppings $t\sim 0.4$ eV ($\sim 0.2$ eV) for 
$e_g$ ($t_{2g}$) orbitals.\cite{Miz96} Higher bands overlap and mix 
up with $d-p$ transitions, though. 

%%%%%%%%%%%%%%%%%%%%%%%%%%%%%%%%%%%%%%%%%%%%%%%%%%%%%%%%%%%%%%%%%%%%%%%%
%%
%%%%%%%%%%%%%%%%%%%%%%%%%%%%%%%%%%%%%%%%%%%%%%%%%%%%%%%%%%%%%%%%%%%%%%%%
In a particular case of vanadates, one has three optical bands $n=1,2,3$ 
arising from the transitions to: 
  (i) a high-spin state $^4\!A_2$ at energy $U-3J_H$,
 (ii) two degenerate low-spin states $^2T_1$ and $^2E$ at $U$, and 
(iii) $^2T_2$ low-spin state at $U+2J_H$.\cite{Kha01} 
Using $\eta=J_H/U$ we parametrize this multiplet structure by:
$R=1/(1-3\eta)$ and $r=1/(1+2\eta)$. In LaVO$_3$ $xy$ orbitals are 
singly occupied,\cite{Mah92} and one obtains a high-spin contribution 
$H_1^{(c)}(ij)$ for a bond ${\langle ij\rangle}$ along $c$ axis:
\begin{equation}
\label{H1c}
H_1^{(c)}=-\frac{1}{3}JR\Big(\vec S_i\!\cdot\!\vec S_j+2\Big) 
\Big(\textstyle{\frac{1}{4}}-\vec \tau_i\!\cdot\!\vec\tau_j\Big),      
\end{equation}
while for a bond in an $(a,b)$ plane:
\begin{equation}
\label{H1ab}
H_1^{(ab)}=-\frac{1}{6}JR\Big(\vec S_i\!\cdot\!\vec S_j+2\Big)
\Big(\textstyle{\frac{1}{4}}-\tau_i^z\tau_j^z\Big). 
\end{equation}
In Eq. (\ref{H1c}) pseudospin operators $\vec\tau_i$ describe 
low-energy dynamics of (initially degenerate) $xz$ and $yz$ orbital 
doublet at site $i$; this dynamics is quenched in $H_1^{(ab)}$
(\ref{H1ab}). 
Here $\frac{1}{3}(\vec S_i\cdot\vec S_j+2)$ is the projection
operator on the {\it high-spin\/} state for $S=1$ spins. The terms 
$H_n^{(c)}(ij)$ for {\it low-spin\/} excitations ($n=2,3$) contain 
instead the spin operator $(1-\vec S_i\cdot\vec S_j)$ (which guarantees 
that these terms vanish for fully polarized spins on a considered bond,
$\langle\vec S_i\cdot\vec S_j\rangle=1$):
\begin{eqnarray}
\label{H23c}
H_2^{(c)}\!\!&=&\!-\frac{1}{12}J \Big(1-\vec S_i\!\cdot\!\vec S_j\Big)
\Big(\textstyle{\frac{7}{4}}-\tau_i^z \tau_j^z - \tau_i^x \tau_j^x
+5\tau_i^y \tau_j^y\Big),                              \nonumber \\
H_3^{(c)}\!\!&=&\!-\frac{1}{4}Jr \Big(1-\vec S_i\!\cdot\!\vec S_j\Big)
\Big(\textstyle{\frac{1}{4}}+\tau_i^z \tau_j^z+\tau_i^x \tau_j^x
-\tau_i^y \tau_j^y\Big),                               \nonumber \\
\end{eqnarray} 
while again the terms $H_n^{(ab)}(ij)$ differ from $H_n^{(c)}(ij)$ 
only by orbital operators: 
\begin{eqnarray}
\label{H23ab}
H_2^{(ab)}\!\!&=&\!-\frac{1}{8}J\Big(1-\vec S_i\!\cdot\!\vec S_j\Big)
\Big(\textstyle{\frac{19}{12}}\mp \textstyle{\frac{1}{2}}\tau_i^z 
\mp \textstyle{\frac{1}{2}}\tau_j^z
-\textstyle{\frac{1}{3}}\tau_i^z\tau_j^z\Big),             \nonumber \\
H_3^{(ab)}\!\!&=&\!-\frac{1}{8}Jr\Big(1-\vec S_i\!\cdot\!\vec S_j\Big)
\Big(\textstyle{\frac{5}{4}}\mp \textstyle{\frac{1}{2}}\tau_i^z 
\mp \textstyle{\frac{1}{2}}\tau_j^z+\tau_i^z\tau_j^z\Big), \nonumber \\         
\end{eqnarray}
where upper (lower) sign corresponds to $a$($b$)-axis bonds. 

%%%%%%%%%%%%%%%%%%%%%%%%%%%%%%%%%%%%%%%%%%%%%%%%%%%%%%%%%%%%%%%%%%%%%%%%
%%
%%%%%%%%%%%%%%%%%%%%%%%%%%%%%%%%%%%%%%%%%%%%%%%%%%%%%%%%%%%%%%%%%%%%%%%%
First we present a mean-field (MF) approximation for the spin and 
orbital bond correlations which are determined self-consistently after 
decoupling them from each other in ${\cal H}_J$ (\ref{sex}). Spin 
interactions,
\begin{equation}
\label{Hspin}
H_s=J_{ab}^s\sum_{\langle ij\rangle_{ab}}\vec S_i\cdot\vec S_j
      -J_c^s\sum_{\langle ij\rangle_c}   \vec S_i\cdot\vec S_j,
\end{equation}
depend on exchange constants: 
\begin{eqnarray}
J_c^s\!\!&=&\!
\frac{1}{2}J\Big[\eta R-(R\!-\!\eta R\!-\!\eta r)
(\textstyle{\frac{1}{4}}+\langle\vec \tau_i\!\cdot\!\vec\tau_j\rangle)
\!-\!2\eta r \langle \tau_i^y \tau_j^y \rangle\Big],     \nonumber  \\
J_{ab}^s\!\!&=&\!\frac{1}{4}J\Big[1\!-\!\eta R\!-\!\eta r+
(R\!-\!\eta R\!-\!\eta r)(\textstyle{\frac{1}{4}}
+\langle\tau_i^z\tau_j^z\rangle)\Big],                  
\label{Jspin}
\end{eqnarray}
determined by orbital correlations. In the orbital sector one finds 
\begin{equation}
\label{Horb}
H_{\tau}=
\sum_{\langle ij\rangle_{c}}\big[J_c^{\tau}\vec\tau_i\cdot\vec\tau_j
-J(1-s_c)\eta r\tau_i^y\tau_j^y\big]
+J_{ab}^{\tau}\sum_{\langle ij\rangle_{ab}}\tau_i^z\tau_j^z,
\end{equation}
with:
\begin{eqnarray}
J_c^{\tau}&=&
\frac{1}{2}J\Big[(1+s_c)R+(1-s_c)\eta(R+r)\Big],        \nonumber  \\
J_{ab}^{\tau}&=&
\frac{1}{4}J\Big[(1-s_{ab})R+(1+s_{ab})\eta(R+r)\Big],  
\label{Jorb}
\end{eqnarray}
depending on spin correlations: 
$s_c=\langle \vec S_i\cdot\vec S_j\rangle_c$ and
$s_{ab}=-\langle\vec S_i\cdot\vec S_j\rangle_{ab}$. In a classical 
$C$-AF state ($s_c=s_{ab}=1$), this MF procedure becomes exact, and 
the orbital problem maps to Heisenberg pseudospin chains along $c$ axis, 
weakly coupled (as $\eta\ll 1$) along $a$ and $b$ bonds,
\begin{equation}
\label{Horb0}
H_{\tau}^{(0)}=JR\Big[\sum_{\langle ij\rangle_c}\vec\tau_i\cdot\vec\tau_j
+\frac{1}{2}\eta\Big(1+\frac{r}{R}\Big)\sum_{\langle ij\rangle_{ab}}
\tau_i^z\tau_j^z\Big],
\end{equation}
releasing large zero-point energy. Thus, spin $C$-AF order and quasi 1D 
quantum orbital fluctuations support each other.\cite{Kha01} 

%%%%%%%%%%%%%%%%%%%%%%%%%%%%%%%%%%%%%%%%%%%%%%%%%%%%%%%%%%%%%%%%%%%%%%%%
%%
%%%%%%%%%%%%%%%%%%%%%%%%%%%%%%%%%%%%%%%%%%%%%%%%%%%%%%%%%%%%%%%%%%%%%%%%
In addition to spin-orbital SE ${\cal H}_J$ (\ref{sex}), orbitally 
degenerate systems experience the Jahn-Teller (JT) interactions --- 
coupling of orbitals to lattice distortions may lead to a structural 
transition, lifting orbital degeneracy by the term, 
\begin{equation}
\label{HV}
{\cal H}_V=V_{ab}\sum_{\langle ij\rangle_{ab}}\tau_i^z\tau_j^z
-V_c\sum_{\langle ij\rangle_c}\tau_i^z\tau_j^z,
\end{equation}
in the present vanadate model.\cite{Kha01} Interactions $V_{ab}>0$ 
originate from the coupling of nearest-neighbor $t_{2g}$ orbitals in 
$(a,b)$ planes to the bond stretching oxygen vibrations in corner-shared 
perovskite structure. They generate antidistortive oxygen displacements 
and staggered orbital order ({\it supporting} SE), whereas the $V_c>0$ 
term due to the GdFeO$_3$-type distortion \cite{Miz99} favors 
ferro-orbital alignment along $c$ axis, and thus {\it competes} with SE. 

%%%%%%%%%%%%%%%%%%%%%%%%%%%%%%%%%%%%%%%%%%%%%%%%%%%%%%%%%%%%%%%%%%%%%%%%
%%
%%%%%%%%%%%%%%%%%%%%%%%%%%%%%%%%%%%%%%%%%%%%%%%%%%%%%%%%%%%%%%%%%%%%%%%%
The complete model ${\cal H}={\cal H}_J+{\cal H}_V$ represents 
a nontrivial many-body problem. Interactions are highly frustrated, 
leading to strong competition between different spin and orbital states. 
We leave this complex problem for a future study, and present here the 
conceptually simpler case of LaVO$_3$ with $C$-AF order. In this phase 
spins are FM along the $c$ axis at low $T$, and orbitals fluctuate on 
their own. This justifies {\it a posteriori\/} a perturbative treatment 
of joint spin-orbital correlations and allows one to determine them in 
a simple analytical way. 

%%%%%%%%%%%%%%%%%%%%%%%%%%%%%%%%%%%%%%%%%%%%%%%%%%%%%%%%%%%%%%%%%%%%%%%%
%%
%%%%%%%%%%%%%%%%%%%%%%%%%%%%%%%%%%%%%%%%%%%%%%%%%%%%%%%%%%%%%%%%%%%%%%%%
We begin with the orbital Hamiltonian $H_{\tau}+{\cal H}_V$ [Eqs. 
(\ref{Horb}) and (\ref{HV})] which has the form of an $XYZ$ model. 
As interchain couplings are weak and of $\tau_i^z\tau_j^z$ form, 
the problem is best handled by employing Jordan-Wigner fermion 
representation.\cite{Mat85} After decoupling $\tau_i^z\tau_j^z$ terms in 
fermionic density and bond-order terms, one finds that the staggered 
orbital order parameter $\tau=|\langle\tau^z_i\rangle|$, the orbital 
ordering temperature $T_{\tau}$, and the temperature dependence of 
orbital correlations:
$\langle\tau_i^{x(y)}\tau_j^{x(y)}\rangle_c=-\frac{1}{2}\kappa$,
$\langle\tau_i^z\tau_j^z\rangle_c=-\tau^2-\kappa^2$,
$\langle\tau_i^z\tau_j^z\rangle_{ab}=-\tau^2$, 
do follow from two self-consistent equations:
\begin{eqnarray}
\label{tau}
\tau&=&\sum_k \Big(\frac{h^{\tau}}{2\varepsilon_k}\Big) 
\;\tanh \Big(\frac{\varepsilon_k}{2T}\Big),                \\
\label{kappa}
\kappa&=&\sum_k \Big(\frac{\tilde J_c}{2\varepsilon_k}\Big) 
\;\cos^2\!k \;\tanh \Big(\frac{\varepsilon_k}{2T}\Big), 
\end{eqnarray}
where $\varepsilon_k=[\tilde J_c^2 \cos^2\!k+(h^{\tau})^2]^{1/2}$ is the 
1D orbiton dispersion, $\tilde J_c=J_c^{\tau}+2\kappa (J_c^{\tau}-V_c)$,  
and $h^{\tau}=2\tau (J_c^{\tau}+2J_{ab}^{\tau}+2V_{ab}-V_c)$ is the 
effective field. We set $k_B=1$.

%%%%%%%%%%%%%%%%%%%%%%%%%%%%%%%%%%%%%%%%%%%%%%%%%%%%%%%%%%%%%%%%%%%%%%%%
%%
%%%%%%%%%%%%%%%%%%%%%%%%%%%%%%%%%%%%%%%%%%%%%%%%%%%%%%%%%%%%%%%%%%%%%%%%
The short-range spin correlations $s_{\gamma}$ determine 
$J_{\gamma}^{\tau}$ (\ref{Jorb}) and are finite also above $T_N$. We 
derived them by solving exactly a single bond $\langle ij\rangle$ within 
the mean-field $\propto\langle S^z\rangle$,\cite{notebf} originating 
from neighboring spins. This is the simplest cluster mean-field theory 
known in the theory of magnetism as Oguchi method.\cite{Ogu60} For a FM 
bond along $c$ axis one finds even an analytic solution:
$s_c\!=\!(Z_0\!-\!Z_1\!-\!2Z_2)/Z$, where 
$Z_0\!=\!1\!+\!2\cosh x\!+\!2\cosh 2x$, 
$Z_1\!=\!(1+2\cosh x)\exp(-2J_c^s/T)$,
$Z_2=\exp(-3J_c^s/T)$, with $Z\!=\!Z_0\!+\!Z_1\!+\!Z_2$, $x=h^s/T$,
$h^s\!=\!(J_c^s+4J_{ab}^s)\langle S^z\rangle$. The $s_{ab}$ correlation
function for an AF bond can be found numerically. 

%%%%%%%%%%%%%%%%%%%%%%%%%%%%%%%%%%%%%%%%%%%%%%%%%%%%%%%%%%%%%%%%%%%%%%%%
%%
%%%%%%%%%%%%%%%%%%%%%%%%%%%%%%%%%%%%%%%%%%%%%%%%%%%%%%%%%%%%%%%%%%%%%%%%
Now we turn to the dynamical coupling between orbital and spin sectors,
denoted as $H_{s\tau}$ term in Eq. (\ref{sex}). In the present case of
$C$-AF ground state it contains mainly contributions due to $c$ axis 
bonds, and reads as follows: 
\begin{equation}
\label{Hstau}
H_{s\tau}\simeq K\sum_{\langle ij\rangle_c}
\delta(\vec S_i\cdot\vec S_j)\delta(\vec\tau_i\cdot\vec\tau_j). 
\end{equation}
Here, $K=\frac{1}{2}J(R-\eta R-\eta r)$ and 
$\delta(A)=A-\langle A\rangle$ implies the fluctuating part of an 
operator, which goes beyond the MF decoupling. Treating 
$H_{s\tau}$ within the high-temperature expansion, we found that 
the joint {\it spin-and-orbital correlations\/}, $f_{ij}=\langle\delta
(\vec S_i\cdot\vec S_j)\delta(\vec\tau_i\cdot\vec\tau_j)\rangle$, 
are given as follows: 
\begin{equation}
f_{ij}=-\frac{3K}{16T}\Big(\langle(\vec S_i\cdot\vec S_j)^2\rangle
-\langle\vec S_i\cdot\vec S_j\rangle^2\Big),
\label{soc}
\end{equation}
with $\langle(\vec S_i\cdot\vec S_j)^2\rangle=1+3Z_2/Z$ in the present 
Oguchi approximation. This high-temperature expansion for $f_{ij}$ is 
valid when spin fluctuations are weak as in the $C$-AF phase, and one 
finds that the joint correlations $f_{ij}$ vanish at $T\to 0$ when fully 
polarized spins decouple from the orbital sector.
 
%%%%%%%%%%%%%%%%%%%%%%%%%%%%%%%%%%%%%%%%%%%%%%%%%%%%%%%%%%%%%%%%%%%%%%%%
%%
%%%%%%%%%%%%%%%%%%%%%%%%%%%%%%%%%%%%%%%%%%%%%%%%%%%%%%%%%%%%%%%%%%%%%%%%
\begin{figure}
\includegraphics[width=8.2cm]{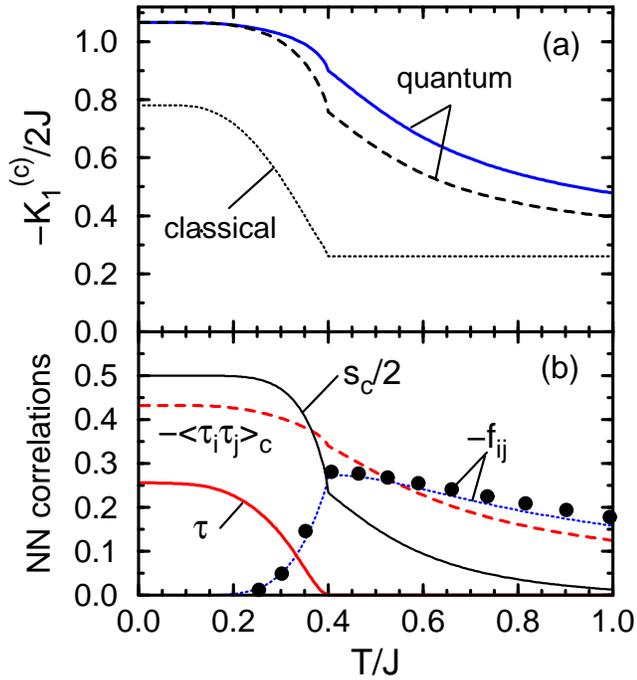}
\caption{(color online)
(a) Kinetic energy $K_1^{(c)}$ for the high-spin excitations 
within classical and quantum models. Solid (dashed) line with (without) 
joint spin-orbital fluctuations.
(b) Intersite correlations along $c$ axis: spin 
$s_c=\langle\vec S_i\!\cdot\!\vec S_j\rangle_c$, orbital 
$\langle\vec\tau_i\cdot\vec\tau_j\rangle_c$, and spin-orbital 
$f_{ij}$. Orbital order parameter $\tau$ is also shown. Dotted line 
(filled circles) for $-f_{ij}$ obtained from high-temperature expansion  
(\ref{soc}) (by exact diagonalization). 
Parameters: $\eta=0.12$, $V_c=0.9J$, $V_{ab}=0.2J$.
}
\label{chs}
\end{figure}

\section{Numerical results}

%%%%%%%%%%%%%%%%%%%%%%%%%%%%%%%%%%%%%%%%%%%%%%%%%%%%%%%%%%%%%%%%%%%%%%%%
%%
%%%%%%%%%%%%%%%%%%%%%%%%%%%%%%%%%%%%%%%%%%%%%%%%%%%%%%%%%%%%%%%%%%%%%%%%
Taking the SE energy scale $J\sim 40$ meV,\cite{Ulr03} we set 
$T_N=0.4J$, while $\eta\simeq 0.12$ follows from spectroscopy. 
\cite{Miz96} It is quite natural in $t_{2g}$ systems that order and 
disorder in the spin and orbital sectors support each other,
\cite{Kha01,Mot03} and indeed $T_{\tau}\simeq T_N$ in LaVO$_3$.
\cite{Miy02} While the microscopic reasons of this behavior are subtle, 
we control $T_{\tau}$ by tuning lattice mediated couplings $V_{\gamma}$ 
--- we found that the orbital transition occurs near $T_N$ for: 
$V_c=0.9J$ and $V_{ab}=0.2J$. Then the energy of the $C$-AF phase is 
still lower than that of the $G$-AF phase, but a slight increase of 
$V_{\gamma}$ (by a factor 1.3) due to a stronger tilting of VO$_6$
octahedra gives the $G$-AF phase, observed in YVO$_3$.\cite{Mah92}
 
%%%%%%%%%%%%%%%%%%%%%%%%%%%%%%%%%%%%%%%%%%%%%%%%%%%%%%%%%%%%%%%%%%%%%%%%
%%
%%%%%%%%%%%%%%%%%%%%%%%%%%%%%%%%%%%%%%%%%%%%%%%%%%%%%%%%%%%%%%%%%%%%%%%%
Following Eqs. (\ref{Jspin}), (\ref{Jorb}), (\ref{tau}), (\ref{kappa}), 
and (\ref{soc}), we calculated orbital, spin, and joint spin-and-orbital 
correlations, and next used them to determine the kinetic energies 
$K_n^{(\gamma)}$ (\ref{central}) due to each Hubbard subband. For
instance,
\begin{equation}
\label{k1}
-K_1^{(c)}/2J=\frac{1}{3}R\Big[
\Big\langle\vec S_i\cdot\vec S_j+2\Big\rangle 
\Big\langle\textstyle{\frac{1}{4}}-\vec\tau_i\cdot\vec\tau_j\Big\rangle
-f_{ij}\Big].      
\end{equation}
Quantum effects beyond the MF theory are particularly pronounced in 
$-K_1^{(c)}$ [see Fig. \ref{chs}(a)]. In the classical approach 
$-K_1^{(c)}$ (\ref{k1}) increases with decreasing $T$ only if $T<T_N$ 
(when both $\tau>0$ and $\langle S^z\rangle>0$). This is qualitatively 
different in the quantum model when the orbitals fluctuate, the orbital 
order parameter $\tau$ is {\it not more than half\/} of its classical 
value, and the spin correlations $s_c$ are finite above $T_N$ and decay 
slowly for $T>T_N$ [Fig. \ref{chs}(b)]. In this case $-K_1^{(c)}$ is 
enhanced at $T=0$ by the orbital fluctuations 
$\langle\vec\tau_i\cdot\vec\tau_j\rangle_c\simeq -0.43$, being close 
to those found for a 1D AF spin $1/2$ chain. Kinetic energy gain [Eq. 
(\ref{k1})] gradually decreases with increasing $T$, and is reduced by 
half at $T\sim 2T_N$ from its value at $T=0$ [Fig. \ref{chs}(a)]. 
We also note that joint spin-and-orbital fluctuations $f_{ij}$ develop 
at finite $T$ and contribute significantly at $T>T_N$.\cite{noteed} An 
opposite behavior occurs in the limit of large JT orbital splitting that 
freezes out orbital fluctuations \cite{Mot03} --- then the temperature 
variation and the anisotropy of optical absorption are quenched, and
$\langle\vec\tau_i\cdot\vec\tau_j\rangle_{\gamma}$ approaches its 
classical limit ($-0.25$) in all directions. 

%%%%%%%%%%%%%%%%%%%%%%%%%%%%%%%%%%%%%%%%%%%%%%%%%%%%%%%%%%%%%%%%%%%%%%%%
%%
%%%%%%%%%%%%%%%%%%%%%%%%%%%%%%%%%%%%%%%%%%%%%%%%%%%%%%%%%%%%%%%%%%%%%%%%
\begin{figure}
\includegraphics[width=8.7cm]{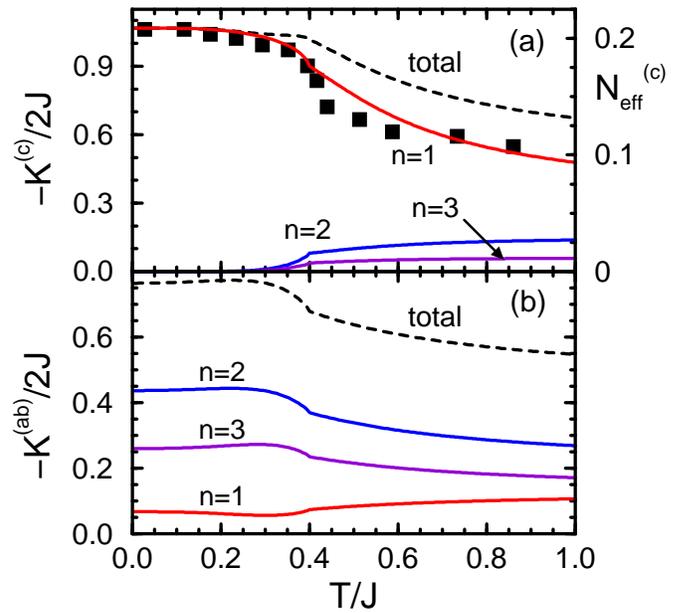}
\caption{(color online)
Kinetic energy $K_n^{(\gamma)}$ (solid lines) and total $K^{(\gamma)}$
(dashed lines) in units of $2J$ for: 
(a) $c$ axis and (b) $ab$-plane polarization. Parameters as in Fig. 1. 
Filled squares in panel (a) represent the effective carrier number 
$N_{\rm eff}^{(c)}$ in LaVO$_3$ which includes the sum of the peaks 1 
and 2 below 3 eV in Fig. 3 of Ref.~\onlinecite{Miy02}. The 
experimental data follows well the calculated intensity of the 
high-spin transition, $n=1$.}
\label{weights}
\end{figure}

%%%%%%%%%%%%%%%%%%%%%%%%%%%%%%%%%%%%%%%%%%%%%%%%%%%%%%%%%%%%%%%%%%%%%%%%
%%
%%%%%%%%%%%%%%%%%%%%%%%%%%%%%%%%%%%%%%%%%%%%%%%%%%%%%%%%%%%%%%%%%%%%%%%%
The optical intensities 
$N_{{\rm eff},n}^{(\gamma)}\propto K_n^{(\gamma)}$ (\ref{sumrule}) 
exhibit pronounced anisotropy between $c$ and $ab$ polarizations (Fig. 
\ref{weights}), particularly those of high-spin transitions ($n=1$) at 
low energy $U-3J_H$. These low-energy 
intensities behave here in opposite way for $c$ and $ab$ polarizations 
when temperature increases and the spectral weight is transferred 
between the high-spin and low-spin bands. The total intensities have a 
much weaker temperature dependence than individual contributions. The 
theory reproduces quite well the observed\cite{Miy02} variation of 
$N_{\rm eff}^{(c)}$ with $T$ [Fig. \ref{weights}(a)]. We recall that
the temperature variation of the kinetic energy $K_1^{(c)}$ and  
$N_{\rm eff}^{(c)}$ is due to evolution of {\it both spin and orbital}
correlations; spin-only ordering with frozen orbitals cannot explain 
a factor of two enhancement of $N_{\rm eff}^{(c)}$ in a range of 
$T<2T_N$. Therefore, the optical data of Ref. \onlinecite{Miy02} 
indicate that the JT orbital splitting in LaVO$_3$ cannot be large.
\cite{notemo} 

\begin{figure}
\includegraphics[width=8.2cm]{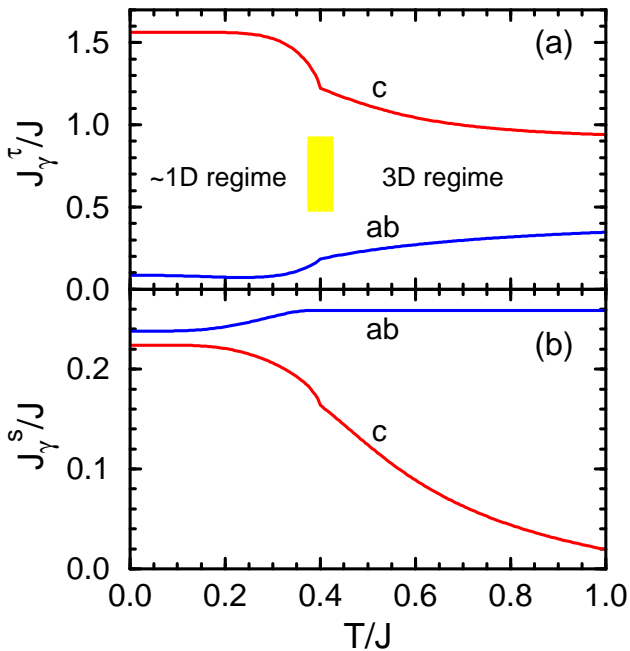}
\caption{(color online)
Exchange constants as functions of $T$ along $c$ 
($a/b$) axis: (a) orbital $J_{\gamma}^{\tau}$ (\ref{Jorb}), and 
(b) spin $J_{\gamma}^s$ (\ref{Jspin}). Parameters: $\eta=0.12$,
$V_c=0.9J$, $V_{ab}=0.2J$.
}
\label{allj}
\end{figure}

%%%%%%%%%%%%%%%%%%%%%%%%%%%%%%%%%%%%%%%%%%%%%%%%%%%%%%%%%%%%%%%%%%%%%%%%
%%
%%%%%%%%%%%%%%%%%%%%%%%%%%%%%%%%%%%%%%%%%%%%%%%%%%%%%%%%%%%%%%%%%%%%%%%%
The microscopic reasons of anisotropy in the optical absorption are
revealed by studying the effective exchange constants, given by Eqs. 
(\ref{Jspin}) and (\ref{Jorb}). While the spin sector is always 3D, 
the orbital one shows a {\it dimensional crossover\/} from 3D to quasi 
1D correlations when the $C$-AF order develops [Fig. \ref{allj}(a)]. 
The orbital singlet correlations along the $c$ axis enhance strongly 
FM $J_{c}^s$ at low $T$. Unlike in pure spin systems, the exchange 
interactions $J_{\gamma}^s$ are {\it temperature dependent\/} [Fig. 
\ref{allj}(b)], and $J_{c}^s$ decreases fast with decreasing intersite 
orbital correlations [Fig. \ref{chs}(b)]. Only at $T\gg T_N$ it is much 
weaker than the AF one, in agreement with conventional 
Goodenough-Kanamori rules. 

%%%%%%%%%%%%%%%%%%%%%%%%%%%%%%%%%%%%%%%%%%%%%%%%%%%%%%%%%%%%%%%%%%%%%%%%
%%
%%%%%%%%%%%%%%%%%%%%%%%%%%%%%%%%%%%%%%%%%%%%%%%%%%%%%%%%%%%%%%%%%%%%%%%%

\section{Discussion and summary}

%%%%%%%%%%%%%%%%%%%%%%%%%%%%%%%%%%%%%%%%%%%%%%%%%%%%%%%%%%%%%%%%%%%%%%%%
%%
%%%%%%%%%%%%%%%%%%%%%%%%%%%%%%%%%%%%%%%%%%%%%%%%%%%%%%%%%%%%%%%%%%%%%%%%
The basic experimental findings in the optical spectra of LaVO$_3$, 
\cite{Miy02} such as:
 (i) pronounced temperature dependence of $c$ axis intensity (changing 
     by a factor of two below 300 K), 
(ii) large anisotropy between the optical spectral weights along $c$ 
     and $a/b$ axis (both below and above $T_N$), 
are qualitatively reproduced by our theory (see Fig. \ref{weights}).
\cite{noteso} This strongly supports the picture of quantum orbital 
chains in the $C$-AF phase of vanadates. 

The spectral shape of optical absorption in LaVO$_3$ is highly 
intriguing, developing sharp coherent peak below 3 eV at $T<T_N$.
\cite{Miy02} When its two-peak structure is interpreted\cite{Miy02} as 
following from the multiplet splitting $\sim 3J_H$, the value of $J_H$ 
would be seriously underestimated, giving $J_H\simeq 0.21$ eV, which is 
much smaller that the respective value (Kanamori parameter) $0.68$ eV
in the free ion.\cite{Miz96} Therefore, we suggest that the absorption
below 3 eV is arising entirely from {\it high-spin band\/} centered at 
$U-3J_H$. The coherent peak emerging at low temperature could be then 
interpreted as a bound state, similar to what occurs in 1D systems.
\cite{Neu98} This is quite natural when spins along the $c$ axis are 
fully polarized and coherent orbital chains are formed. 

%%%%%%%%%%%%%%%%%%%%%%%%%%%%%%%%%%%%%%%%%%%%%%%%%%%%%%%%%%%%%%%%%%%%%%%%
%%
%%%%%%%%%%%%%%%%%%%%%%%%%%%%%%%%%%%%%%%%%%%%%%%%%%%%%%%%%%%%%%%%%%%%%%%%
While the precise structure of the optical band requires further work, 
we may already obtain some useful information. Taking the width of $8t$ 
for the optical band,\cite{Neu98} and fitting the position and width 
of the spectral density in Fig. 2(a) of Ref. \onlinecite{Miy02}, one 
finds $U-3J_H\sim 2.3$ eV, $t\sim 0.2$ eV, and thus $JR\sim 70$ meV. As 
higher bands with $n=2,3$ are not resolved in the experiment, we cannot 
fix $J_H$ directly from the multiplet structure, but considering 
$\eta=J_H/U\simeq 0.13$ we reproduce basic 
energy scale $J\simeq 42$ meV, being close to $J\sim 40$ meV deduced 
from neutron scattering data.\cite{Ulr03} In fact, we can also determine 
$J$ directly from the comparison of the theoretical result 
$-K_1^{(c)}\simeq 1.1(2J)$, and the observed 
$N_{\rm eff}^{(c)}\simeq 0.21$ at $T=0$ [see Fig. \ref{weights}(a)]. 
Using Eq. (\ref{sumrule}) with $a_0=3.91$ \AA,\cite{Mah92} one obtains 
then $J\simeq 48$ meV, a value remarkably consistent with both above 
estimates. Finally, we remark that the present interpretation of the 
experimental data of Ref. \onlinecite{Miy02} gives 
(at $\eta\simeq 0.13$) $J_H\simeq 0.5$ 
eV and $U\simeq 3.8$ eV, which are somewhat lower than the atomic 
values.\cite{Miz96} This reduction may be attributed to covalency and/or 
many-body screening effects in a solid.\cite{note05}    

%%%%%%%%%%%%%%%%%%%%%%%%%%%%%%%%%%%%%%%%%%%%%%%%%%%%%%%%%%%%%%%%%%%%%%%%
%%
%%%%%%%%%%%%%%%%%%%%%%%%%%%%%%%%%%%%%%%%%%%%%%%%%%%%%%%%%%%%%%%%%%%%%%%%
Summarizing, we proposed a new approach employing {\it partial optical 
sum rules\/} in Mott insulators with orbital degeneracy, which provides 
a theoretical framework for common understanding of the optical and  
magnetic experiments, both determined by the superexchange. Considering 
the example of LaVO$_3$ with $C$-AF order, we have shown that pronounced 
temperature dependence and strong anisotropy of the optical absorption 
indeed follow from quasi 1D quantum spin-orbital correlations, being 
radically different from the classical expectations. A satisfactory 
agreement between the values of $J$ extracted independently from the 
magnetic and optical data in LaVO$_3$ demonstrates that superexchange 
interactions are indeed responsible for the distribution of the optical 
spectral weight in Mott insulators. 

%%%%%%%%%%%%%%%%%%%%%%%%%%%%%%%%%%%%%%%%%%%%%%%%%%%%%%%%%%%%%%%%%%%%%%%%
%%
%%%%%%%%%%%%%%%%%%%%%%%%%%%%%%%%%%%%%%%%%%%%%%%%%%%%%%%%%%%%%%%%%%%%%%%%
\begin{acknowledgments}
We thank B. Keimer and Y. Tokura for insightful discussions and T. Enss
for his help on the graphical presentation of the experimental data in
Fig. 2. 
A.~M.~Ole\'s would like to acknowledge support by the Polish State 
Committee of Scientific Research (KBN) under Project No.~1~P03B~068~26. 
\end{acknowledgments}

%%%%%%%%%%%%%%%%%%%%%%%%%%%%%%%%%%%%%%%%%%%%%%%%%%%%%%%%%%%%%%%%%%%%%%%%
%%
%%%%%%%%%%%%%%%%%%%%%%%%%%%%%%%%%%%%%%%%%%%%%%%%%%%%%%%%%%%%%%%%%%%%%%%%


\begin{thebibliography}{}

\bibitem{And59} P. W. Anderson, Phys. Rev. {\bf 115}, 2 (1959).

\bibitem{Bae86} D. Baeriswyl, J. Carmelo, and A. Luther, 
                   \prb {\bf 33}, 7247 (1986).

\bibitem{Aic02}	M. Aichhorn, P. Horsch, W. von der Linden, and M.~Cuoco, 
		   \prb {\bf 65}, 201101(R) (2002).

\bibitem{Kug82} K. I. Kugel and D. I. Khomskii,
                   Sov. Phys. Usp. {\bf 25}, 231 (1982);
                Y. Tokura and N. Nagaosa,
                   Science {\bf 288}, 462 (2000);
		Y. Tokura,
		   Physics Today, July 2003, p. 50.   

\bibitem{Tob01} K. Tobe, T. Kimura, Y. Okimoto, and Y. Tokura, 
                   \prb {\bf 64}, 184421 (2001).

\bibitem{noteaaf} $A$-type ($C$-type) AF phase consists of 
                  ferromagnetic planes (chains) with AF order
                  between them. 

\bibitem{Miy02} S. Miyasaka, Y. Okimoto, and Y. Tokura,
                   J. Phys. Soc. Jpn. {\bf 71}, 2086 (2002).   

\bibitem{Mah92} A. V. Mahajan, D. C. Johnston, D. R. Torgeson, 
                   and F. Borsa,
                   \prb {\bf 46}, 10966 (1992);
                S. Miyasaka, T. Okuda, and Y. Tokura,
                   \prl {\bf 85}, 5388 (2000);
                Y. Ren, T. T. M. Palstra, D. I. Khomskii, 
		   A.~A.~Nugroho, A. A. Menovsky, and G. A. Sawatzky,
		   \prb {\bf 62}, 6577 (2000);
                M. Noguchi, A. Nakazawa, S. Oka, T.~Arima, 
		   Y. Wakabayashi, H. Nakao, and Y. Murakami,
                   {\it ibid.\/} {\bf 62}, R9271 (2000);
                G. R. Blake, T. T. M. Palstra, Y. Ren, A. A. Nugroho,
		   and A. A. Menovsky,
                   \prl {\bf 87}, 245501 (2001). 

\bibitem{Ulr03} C. Ulrich, G. Khaliullin, J. Sirker, M. Reehuis, 
                   M. Ohl, S.~Miyasaka, Y. Tokura, and B. Keimer, 
                   \prl {\bf 91}, 257202 (2003).

\bibitem{Kha01} G. Khaliullin, P. Horsch, and A. M. Ole\'{s}, 
                   \prl {\bf 86}, 3879 (2001).

\bibitem{Lee02} J. S. Lee, Y. S. Lee, T. W. Noh, S.-J. Oh, Jaejun Yu, 
                   S.~Nakatsuji, H. Fukazawa, and Y. Maeno,
                   \prl {\bf 89}, 257402 (2002).

\bibitem{Fei99} L. F. Feiner and A. M. Ole\'{s}, 
                   \prb {\bf 59}, 3295 (1999).

\bibitem{Kha00} G. Khaliullin and S. Maekawa, 
                   \prl {\bf 85}, 3950 (2000).

\bibitem{Miz96} T. Mizokawa and A. Fujimori,
                   \prb {\bf 54}, 5368 (1996).

\bibitem{Miz99} T. Mizokawa, D. I. Khomskii, and G. A. Sawatzky,
                   \prb {\bf 60}, 7309 (1999).

\bibitem{Mat85} D. C. Mattis,
                   {\it The Theory of Magnetism II\/}
                   (Springer-Verlag, New York, 1985).

\bibitem{notebf} We assume that the spin order parameter follows the 
                 spin-wave expansion, 
                 $\delta\langle S^z\rangle_T\propto (T/T_N)^2$ at 
		 $T\ll T_N$, and has a MF behavior 
		 $\langle S^z\rangle\propto (1-T/T_N)^{1/2}$ when 
		 $T\to T_N$.

\bibitem{Ogu60} T. Oguchi, 
		   Progr. Theor. Phys. (Kyoto), {\bf 13}, 148 (1955);
                see also: Y. Shibata, S. Nishimoto, and Y. Ohta, 
                   \prb {\bf 64}, 235107 (2001).

\bibitem{Mot03} Y. Motome, H. Seo, Z. Fang, and N. Nagaosa, 
                   \prl {\bf 90}, 146602 (2003). 

\bibitem{noteed} Exact diagonalization of a four-site chain embedded 
                 in self-consistent spin and orbital fields gave very 
                 similar values of $f_{ij}$ to those resulting from 
		 Eq. (\protect\ref{soc}) [Fig. \protect\ref{chs}(b)], 
		 demonstrating a surprisingly good quality of the 
		 high temperature (\protect\ref{soc}) expansion for 
		 $f_{ij}$ in the $C$-AF phase.

\bibitem{notemo} Even taking a small JT splitting, these data cannot 
                 be explained by a pure orbital model of Ref.
                 \protect\onlinecite{Mot03}, and requires 
		 consideration of full spin-orbital problem. 

\bibitem{noteso} A quantitative comparison with experiment would 
                 require also the spin-orbit coupling 
		 $\propto\lambda\vec S_i\cdot\vec l_i$, 
		 which reduces to $\lambda S_i^zl_i^z\!\equiv
		 2\lambda S_i^z\tau_i^y$ when $xy$ orbital is quenched
		 [P. Horsch, G. Khaliullin, and A. M. Ole\'{s},
                     \prl {\bf 91}, 257203 (2003)].
 		 Below $T_N$ this coupling acts as transverse field 
		 applied to the orbital chains, {\it competing\/} with 
		 both ${\cal H}_J$ and ${\cal H}_V$. Possible 
		 fluctuations of $xy$ orbitals, not included in the 
		 present theoretical approach, might be also of some 
		 importance for a more favorable comparison with
		 experiment. 

\bibitem{Neu98} R. Neudert, M. Knupfer, M. S. Golden, J. Fink, 
                   W. Stephan, K. Penc, N. Motoyama, H. Eisaki, 
		   and S.~Uchida, 
                   \prl {\bf 81}, 657 (1998).

\bibitem{note05} For obtained $U$ and $J_H$, the low-spin transitions
                 with $n=2,3$ are expected to be located at $\sim 3.8$
		 eV and $\sim 4.8$ eV, respectively. These transitions
		 loose their intensity below $T_N$ [see Fig.
		 \protect\ref{weights}(a)], as actually seen in the
		 data of Ref. \protect\onlinecite{Miy02}.

\end{thebibliography}
\end{document}